\documentclass[5p]{elsarticle}

\usepackage{hyperref}
\usepackage{amsmath,amssymb,amsfonts}
\usepackage{algorithmic}
\usepackage{graphicx}
\usepackage[british]{babel}
\usepackage{textcomp}
\usepackage{array}
\usepackage{url}
\usepackage{amsbsy}
\usepackage{epstopdf}
\usepackage[table]{xcolor}
\usepackage{multirow}
\usepackage{hhline}
\usepackage{rotating}

\usepackage{caption}
\usepackage{subcaption}
\begin{document}

\begin{frontmatter}

\title{Convolutional Neural Networks for Video Quality Assessment \tnoteref{t1}}
\tnotetext[t1]{This is the author's own write-up of research results and analysis that has not been peer reviewed, nor had any other value added to it by a publisher (such as formatting, copy-editing, technical enhancements, and the like).}


\author[UNI_CRETE,FORTH_ICS]{Michalis Giannopoulos}
\ead{mgiannop@ics.forth.gr}
\author[FORTH_ICS]{Grigorios Tsagkatakis}
\author[BBC]{Saverio Blasi}
\author[QMUL]{Farzad Toutounchi}
\author[FORTH_ICS]{Athanasios Mouchtaris}

\address[UNI_CRETE]{Department of Computer Science, University of Crete, Greece}
\author[FORTH_ICS]{Panagiotis Tsakalides}
\address[FORTH_ICS]{Institute of Computer Science, Foundation for Research and Technology Hellas, Greece}


\author[BBC]{Marta Mrak}
\address[BBC]{BBC Research and Development, UK}

\author[QMUL]{Ebroul Izquierdo}
\address[QMUL]{Queen Mary University of London, UK}

\begin{abstract}

Video Quality Assessment (VQA) is a very challenging task due to its highly subjective nature. Moreover, many factors influence VQA. Compression of video content, while necessary for minimising transmission and storage requirements, introduces distortions which can have detrimental effects on the perceived quality. Especially when dealing with modern video coding standards, it is extremely difficult to model the effects of compression due to the unpredictability of encoding on different content types. Moreover, transmission also introduces delays and other distortion types which affect the perceived quality. Therefore, it would be highly beneficial to accurately predict the perceived quality of video to be distributed over modern content distribution platforms, so that specific actions could be undertaken to maximise the Quality of Experience (QoE) of the users. Traditional VQA techniques based on feature extraction and modelling may not be sufficiently accurate. In this paper, a novel Deep Learning (DL) framework is introduced for effectively predicting VQA of video content delivery mechanisms based on end-to-end feature learning. The proposed framework is based on Convolutional Neural Networks, taking into account compression distortion as well as transmission delays. Training and evaluation of the proposed framework are performed on a user annotated VQA dataset specifically created to undertake this work. The experiments show that the proposed methods can lead to high accuracy of the quality estimation, showcasing the potential of using DL in complex VQA scenarios.
\end{abstract}


\end{frontmatter}


\section{Introduction} \label{First_Section}
Due to the large size of uncompressed video signals, video compression is essential to ensure that content can be efficiently and timely distributed to final users. In order to obtain the low bit-rates required for smooth video delivery over conventional networks, video distribution systems rely on lossy compression. Modern video coding solutions such as the H.265/High Efficiency Video Coding (HEVC) standard \cite{HEVC} are capable of achieving very high compression ratios, while minimising the distortion introduced during compression. Nonetheless, the processing performed by HEVC encoders may introduce some amount of distortion or artefacts in the video signal, which negatively impact the user's perceived quality of the received video signal.

HEVC relies on a flexible approach which allows content to be compressed with different strengths (defined by the Quantisation Parameter, QP, used within the compression loop), depending on the application. Higher QPs result in lower quality of the compressed signal, at smaller average bit-rates, whereas low QPs result in better quality content at higher bit-rates. When distributing content in challenging network conditions, it is crucial to ensure that bit-rates are not too high and do not exceed the capacity of the network, otherwise users will experience delays which also contribute negatively to the perceived quality of the received signal. Moreover, it is worth noticing that using the same settings on different video content may produce very different results in terms of quality and bit-rate of the compressed signal. This is due to the complex mechanisms adopted by standards such as HEVC, which rely on exploiting spatial, temporal and statistical redundancies within the video signal to achieve compression. Due to the unpredictability of the compression step, it is very difficult to predict the Quality of Experience (QoE) perceived by viewers watching compressed content distributed through a transmission network.

QoE is defined by the International Telecommunication Union (ITU) as "the overall acceptability of an application or service, as perceived subjectively by the end-user" \cite{rec2007p}. As such, QoE may include the complete end-to-end system effects (acquisition, processing, compression, storage, transmission, etc.), as well as being influenced by user expectations and context. In this paper, focus is given to perceived subjective Video Quality Assessment (VQA), taking into account the effects of coding distortion and transmission delays. Subjective VQA can be measured in psychophysical experiments in which a number of subjects rate a given set of content. Depending on the application, tests can be performed with Full-Reference (FR), where viewers are asked to compare the processed video against the original reference; or tests can be performed with Reduced-Reference, in which the comparison happens based on a specific number of features from the reference video; or finally, in case only the processed videos are presented to the viewers, No-Reference (NR) tests can be performed. The latter is the case for the NR subjective tests used within this paper (which will be described in more details in the rest of the paper), in that the assessment was performed at the receiving end, where there is no availability of the uncompressed signal before transmission.

In order to obtain representative ratings, a certain number of non-expert viewers (to avoid potential subject bias) should be invited. According to ITU-T (recommendation P.910) \cite{itu2008p910}, any number between $4$ and $40$ is desirable. Moreover, due to possible influence factors from heterogeneous contexts, tests should be performed in a neutral environment (e.g. a dedicated laboratory room). After the tests, the scores for all participants are averaged to compute so-called Mean Opinion Scores (MOS). Obviously, preparing and running such tests can be expensive and time consuming.

For that reason, methods to objectively predict the VQA of video content are highly desirable. In this paper, a Deep Learning (DL) approach for automatic VQA based on Convolutional Neural Networks (CNN) is proposed, with the goal of predicting the expected perceived quality of compressed video content after transmission. The system is tailored for very challenging applications entailing usage of User Generated Content (UGC) which is more and more relevant in many scenarios. As such, the proposed system was trained and tested under challenging conditions both from the compression perspective (in that content contains noise, fast motion, and is in general of lower quality than professionally-captured video content) as well as the transmission (in that mobile device users may be in areas with low network coverage).

The system is capable of making the VQA prediction at the source and, differently than other methods, it accepts as input raw visual data, without performing any further processing or transmission, thus reducing the necessary complexity. Moreover, differently than other alternative techniques, the supervised DL approach assures end-to-end feature learning, and the regression-flavoured task is transformed into a classification task, aiming to provide results which are easily exploitable within the distribution chain, as illustrated in the rest of this paper. 

The rest of this paper is organised as follows. Section \ref{Second_Section} briefly summarises the basic concepts of existing approaches towards NR VQA methods and metrics, provides basic notations and preliminaries of supervised DL prediction techniques, and gives an overview of state-of-the-art methods. In Section \ref{Third_Section}, the proposed DL-based QoE predictive NR framework is presented, while the evaluation methodology and the respective experiments take place in Section \ref{Fourth_Section}. Section \ref{Fifth_Section} provides extensive experimental evaluation of the proposed model and analysis of the results. Finally, concluding remarks are pointed out in Section \ref{Sixth_Section}.

\section{State of the art}
\label{Second_Section}
The simplest way of evaluating quality of video signals is using FR metrics such as the Peak Signal-to-Noise Ratio (PSNR) \cite{winkler2008evolution}, which is a function of the Mean Square Error (MSE) between each frame of the reference and the processed video signal. PSNR is widely used in video coding for instance for rate-distortion optimisation, where it has proven to work well while being inexpensive to compute. On the other hand, PSNR may not match well with perceived visual quality due to the complex, highly non-linear behaviour of the human visual system \cite{wang2009mean}, and it cannot be used to measure some of the effects of transmission (such as delays), because it does not generalise to the temporal dimension. Similarly, the popular Structural Similarity (SSIM) index \cite{wang2004image}, \cite{wang2004video} is frequently used for estimating video quality. The computation is also performed in a frame-by-frame manner to the luminance component of the video sequence, and, in conjunction with the contrast and structure components, the overall degradation is computed as the average of the SSIM indexes at each frame level. When dealing with variations across scales, SSIM can be extended to Multi-Scale Structural Similarity (MS-SSIM) \cite{wang2003multiscale}. More complex FR metrics have been proposed, including the Video Quality Metric (VQM) \cite{pinson2004new} and the Motion-based Video Integrity Evaluation (MOVIE) \cite{seshadrinathan2010motion}, or the Visual Information Fidelity (VIF) \cite{li2016toward}.

On the other hand, NR metrics have also been proposed to estimate the quality of video content \cite{shahid2014no} \cite{bovik2013automatic}. These metrics typically require lower computational complexity  \cite{yang2007perceptual}, \cite{yang2005novel}, \cite{kawayoke2008nr}, \cite{brandao2010no}, in order to be used on-line for quantifying the quality of video content. A DCT-based approach for estimating the effects on quality of various types of compression distortions is proposed in \cite{saad2014blind}, while a more general approach relying on statistical properties of undistorted videos is presented in \cite{mittal2016completely}. In an overview study \cite{torres2016experimental} a comparison of different metrics is presented, stating that there is no universally effective metric, indicating that for many applications automatic VQA is an open research question.

In addition to the aforementioned models which try to directly model the distortion in a picture, statistical models can also be defined, in which independent variables are fit against results obtained in a subjective quality evaluation test using regression techniques. These methods rely on the availability of a set of pre-annotated training data, generated by means of subjective VQA. As such, instead of directly trying to model the distortions, these methods try to correlate the annotated assessment with the reference signal. One way of achieving this goal is by manually defining specific features that are assumed to be relevant to the subjective quality, and subsequently use a mapping between the feature space and the subjective quality space. Unfortunately, manually designing such mapping may be difficult, and as such these methods may not be ideal especially in cases where the processing consists in complex unpredictable operations, such as those entailed by modern video encoders.

As a possible alternative, ML based methods have been recently proposed. These methods typically rely on two steps: Feature extraction, in which representative features of the video content are computed; and classification, where the extracted features are mapped into class scores based on a trained algorithm. From an abstract perspective, ML schemes usually perform a dimensionality reduction technique to reduce the original data space, followed by a prediction scheme performed by trainable algorithmic methods. Among such methods, Support Vector Machines (SVMs), k-Nearest Neighbors (k-NN) and Decision Trees (DT) have all been used in VQA \cite{mittal2012no}, \cite{le2006convolutional}, \cite{saad2014blind}, \cite{narwaria2012svd}, \cite{plakia2016user}. A typical ML system tries to learn and gain knowledge from the training data it is provided with, in order to be able to make predictions concerning new test data it will be shown. Nonetheless, several issues must be taken into consideration. First of all, ML learning methods depend heavily on sophisticated feature extraction methods designed specifically for a certain task. These methods are based on the assumption that the selected features are relevant to the subjective quality, but varying datasets of the same nature can limit the effectiveness of such assumption. Furthermore, such techniques only consider the distortions introduced by compression, and do not take into account transmission.

A few approaches have been presented dealing with predicting the perceived quality after transmitting video content through a network. In \cite{bampis2017learning}, the authors consider video impairments based on playback interruptions, mainly caused by bandwidth limitations. A set of features is derived, which are then used to fit a regression-based predictive approach \cite{bampis2017study}.  Regression is performed using different models such as Support Vector Regression (SVR), Random Forests (RF) or Gaussian Boosting (GB). In \cite{kumar2015intelligent}, a ML scheme is adopted for wireless communication applications. The proposed method involves a Pseudo-Subjective Quality Assessment (PSQA) procedure, during which a finite set of high-influential parameters is selected, and subsequently video content is rated by subjective viewers. The subjective data is then fed to a regression model. 

In \cite{sogaard2015video} the authors propose a regularised linear regression NR model called Elastic Net (EN). The video features extracted are based on an approach described in \cite{sogaard2015no}, with the goal of estimating the QP used during the encoding and the corresponding  PSNR, similar to the work presented in \cite{bi2003dimensionality}. Another approach is proposed in \cite{rehman2015display}, where the authors present the results of a subjective study in order to assess the effects of viewing conditions and display devices on the VQA process. Furthermore, they propose a FR metric called SSIMplus \cite{ssimplus} which operates in real-time for predicting the quality of video content.

A NR machine learning based approach for streaming applications can be found in \cite{vega2017predictive}. The authors extract eight NR video features (occurring in bit-stream and pixel level) and combine them with the nominal bit-rate and estimated level of packet loss in order to form a representative feature set. This feature set is subsequently fed to regression based predictive algorithms, which carry out the QoE assessment. Such algorithms vary between Multiple Linear Regression (MLR) and Standard Regression Trees to GP regression and SVRs. Finally, a deep unsupervised learning scheme is proposed in \cite{vega2017deep}. In this work, the authors employ Restricted Boltzmann Machines combined with eight NR features.

Inspired by the fact that DL techniques are effective on many problems in image and video processing (for instance image/video classification, human activity recognition, etc.) compared to conventional machine learning techniques, a DL framework for efficient VQA prediction is proposed in this paper. The choice of a DL approach towards VQA is two-folded. Firstly, DL models can acquire remarkable generalisation capabilities when sufficient data is used for training, especially if using data augmentation techniques such as those utilised in the proposed methodology. Secondly, DL models do not depend on sophisticated feature extraction and selection techniques, which is the case with traditional machine learning techniques mentioned above, but perform end-to-end learning and optimisation via linear and non-linear transformations to raw pixel data. Due to the subjective nature of this problem, defining a set of features that appropriately correlate with the final VQA is not trivial, and as such DL approaches are suitable for this task in that they overcome this step. DL models based on Convolutional Neural Networks (CNNs) have recently been used for picture-quality prediction \cite{kim2017deep}. The approach proposed in this paper goes beyond state of the art, by investigating usage of CNNs towards higher-order models. In particular, temporal information is considered by feeding three dimensional patches to the algorithm, ensuring that variations of quality over time are taken into account. The VQA is posed as a classification problem as shown in the following section.

\section{Problem formulation}
\label{Third_Section}
In this section, some background is provided to formulate the problem of predicting VQA using deep neural networks. In addition, a description of the dataset used for training and testing of the approach is provided. The dataset was created specifically to develop the work presented in this paper.

\subsection {VQA as a classification problem}
When applying ML for predictive VQA modelling purposes, training data is used to make a prediction. This prediction should generalise well on new data on which there is no ground truth. As such, predictive modelling can be described as the goal of approximating a mapping function from input variables to output variables. In the case described in this paper, the output variable is a score of the predicted VQA of the current piece of content being considered. The output variable to represent the VQA can be treated as either a discrete or continuous parameter. In the latter case, a continuous scale could be used to score the VQA from a minimum to a maximum. 

In ideal conditions assuming availability of a large training set, this case could better fit the ground truth obtained from subjective testing, which is obtained as the average of the MOS scores provided by the test participants. Such averages are expressed as continuous variables. On the other hand, posing the problem in such a way that the output variable is continuous does not work well for the cases in which limited items are available for training. Under these conditions, the networks have limited ability of learning from the data, and generalisation is more difficult to obtain. For that reason, a discretised output variable can be more suitable, leading to higher accuracy of the prediction, despite the limited size of the training set. In case the output variable is a discrete variable, the task is typically referred to as a classification problem. The input variables in a classification problem can be either real-valued or discrete-valued.

Following from the aforementioned observations, the following scheme is proposed in this paper. The objective is to predict the VQA of a given piece of compressed video content that needs to be transmitted through a transmission network under known network conditions.  The CNN is input raw pixel data. This data is obtained by means of decoding the compressed bitstream, and then pre-processing the data in order to embed the effects of the network within the signal. In this respect, several parameters which affect the network conditions need to be taken into account, including Maximum Segment Size (MMS),Round-Trip Time (RTT), and loss rate. The rate throughput is then estimated (assuming that Transmission Control Protocol, TCP, is used) as:
\begin{equation} \label{rate_equation}
R=\frac{1.22M}{T\sqrt{L}},
\end{equation}
\noindent where $M$ is the MMS, $T$ is the RTT and $L$ is the loss rate. 

Finally, for a given video content with a bit-rate $B$ of length $N$ (in seconds), the following delay can be considered:
\begin{equation} \label{delay_equation}
\delta=\frac{B}{R}N
\end{equation}

\begin{figure}[!ht]
\centering
\includegraphics[width=250pt]{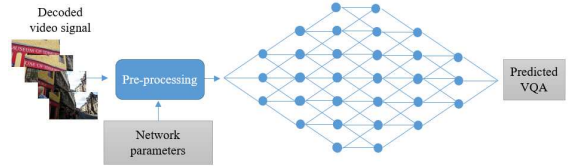}
\caption{Proposed CNN-based approach for VQA prediction.}
\label{Proposed_Method_Compressed_Video_First_Figure}
\end{figure}

The delay $\delta$ is manually added at the beginning of the raw pixel data signal before this is fed to the CNN. Training is performed using a dataset consisting of annotated content (as described in the following of this section). The ground truths used during the training are average MOS training values, which are discretised into a fixed number of classes. In order to estimate the performance of the proposed classification predictive model, the  accuracy of the prediction can be computed, corresponding to the percentage of correctly classified samples over the total number of estimations in a test set. A scheme of the proposed approach is illustrated in Figure \ref{Proposed_Method_Compressed_Video_First_Figure}. 

\subsection{Dataset}
In designing DL solutions and training ML models, the definition of a suitable training and test set is a critical task. With regards to VQA, a set of video sequences with different levels of perceived quality is required. In addition, the annotated  associated perceived QoE for each video is also essential to act as the ground truth in the training process. The subjects should annotate video clips under known conditions in terms of compression parameter as well as network conditions under which the video content was transmitted.

Given the nature of the problem presented in this paper and the general approach depicted in Figure \ref{Proposed_Method_Compressed_Video_First_Figure}, a suitable dataset was difficult to identify in the literature. Therefore, a new dataset was built specifically for the problem at hand. The dataset comprises video clips encoded with different QPs which are transmitted through simulated enforced network conditions. As already mentioned, the method should be capable of dealing with challenging scenarios such as those imposed by using UGC, which is typically of poorer quality than broadcast content.

\begin{figure}[t]
\centering
\includegraphics[width=250pt]{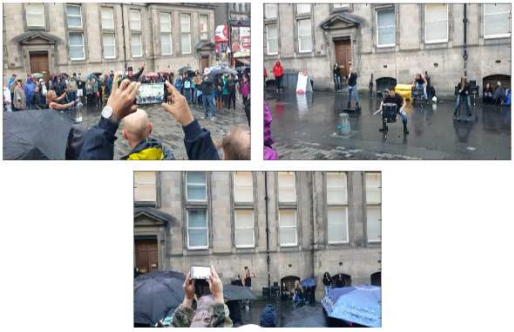}
\caption{The UGC sequences used for the creation of the annotated VQA dataset.}
\label{screenshots}
\end{figure}

In this regard, $3$ UGC video sequences from the open access $2016$ Edinburgh Festival dataset \cite{weerakkody2017} were selected. Each video clip is of $10$ seconds in length, comprising $300$ frames, a spatial resolution of $1920 \times 1080$ samples, and frame rate of $25$ Hz. Exemplary frames from the $3$ aforementioned UGC video sequences can be seen in Figure \ref{screenshots}.

The sequences were encoded with $5$ different QPs. The encoding was performed using a practical HEVC encoder solution. The Turing codec was used at this purpose \cite{Turing}, an open-source practical HEVC software implementation which is capable of compressing video content at very high compression efficiency, while at the same time providing features typical of practical encoders, such as low complexity, high parallelisation capabilities, and minimal memory requirements. This is beneficial to ensure that the proposed approach can be used within practical use case scenarios.

\begin{figure}[!b]
\centering
\begin{subfigure}[!b]{0.5\textwidth}
\includegraphics[width=250pt,height=75pt]{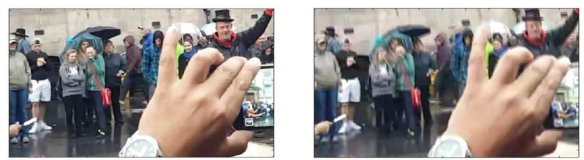}
\caption{UGC sequence 1}
\label{Proposed_Method_Dataset_Description_Second_Figure_UGC1}
\end{subfigure}
\begin{subfigure}[!b]{0.5\textwidth}
\includegraphics[width=250pt,height=74pt]{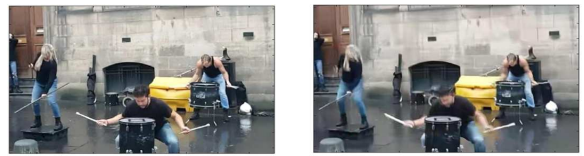}
\caption{UGC sequence 2}
\label{Proposed_Method_Dataset_Description_Second_Figure_UGC2}
\end{subfigure}
\caption{Example frames extracted from two of the UGC sequences used for the dataset, compressed with QP=$18$ (left) and QP=$40$ (right).}
\vspace{-5pt}
\label{Proposed_Method_Dataset_Description_Second_Figure}
\end{figure}

Encoding was performed with different QP values aimed at simulating various compression distortions witnessed in distribution conditions. Hence the values $18$, $24$, $28$, $32$, and $40$ were selected, to cover a wide range of delivery requirements. As already mentioned, the effects of encoding using different QP values on the actual quality of the decoded signal are difficult to predict, and highly content-dependent. As an example, Figure \ref{Proposed_Method_Dataset_Description_Second_Figure} shows frames extracted from two of the three UGC sequences in the dataset, encoded with different QP values. As can be seen, the effects of high quantisation are generally very evident in that the sequence is on average of much lower quality. On the other hand, such effects are not uniform. Smoother areas of content, such as the hand or the background in the sequence shown in Figure \ref{Proposed_Method_Dataset_Description_Second_Figure_UGC1} tend to be compressed more easily even when using higher QP values. Conversely, textured areas with higher amount of details suffer more from usage of high QP values, as is evident from the sequence shown in Figure \ref{Proposed_Method_Dataset_Description_Second_Figure_UGC2}.

In addition to considering the effects of compression, transmission is also considered in the dataset. By considering various combinations of the aforementioned network parameters, four realistic conditions were modelled each represented by a final overall network throughput rate (and corresponding delay $\delta$, obtained as in Equation \ref{delay_equation}). The selected network conditions are listed as the following:
\begin{itemize}
\item $R = 1.70$ Mbps rate associated with $T=50$ms, and $L=3\%$
\item $R = 7.32$ Mbps rate associated with $T=20$ms, and $L=2\%$
\item $R = 20.70$ Mbps rate associated with $T=5$ms, and $L=1\%$
\item $R = 92.60$ Mbps rate associated with $T=5$ms, and $L=0.1\%$
\end{itemize}

An MSS of $M = 1500$ Bytes was considered for all cases. The effects of the above network conditions was incorporated in the encoded sequences by means of pre-processing (as illustrated in Figure \ref{Proposed_Method_Compressed_Video_First_Figure}).

The combination of the different encoding parameters and network parameters resulted in $3\times5\times4=60$ different video clips, that created the basis for the dataset. The video clips and the simulations described above created the training data to be fed to the DL model as the input. During the training phase, the network creates QoE ratings associated with each input that need to be compared with ground truth values. Hence it is necessary to have an annotated dataset of VQA values associated with the described video content. To address this, a subjective assessment of the  video dataset was performed, and the quality of the videos were rated based on the existing distortions introduced by compression and network conditions.

The subjective assessment was based on the standard Absolute Category Rating (ACR) metric recommended by ITU \cite{itu2008p910,series2012methodology}, that measures the perceived quality based on source stimuli that are presented to the viewers separately and are rated independently. The VQA rates are ranged from $1$ to $5$, that represent bad, poor, fair, good, and excellent QoE.

In this regard, $8$ subjects were selected to participate in the subjective assessment. They were asked to watch the $60$ items in the dataset. Viewing conditions were carefully controlled to simulate normal scenarios in which viewers usually watch TV. Normal displays with a native resolution of $1920 \times 1080$ pixels were used for the assessment. Participants were instructed to consider the general quality of the items, considering all the existing conditions in the videos, including the content, visual artefacts and distortions, starting delay, camera movements, etc., giving a single rating from $1$ to $5$ to describe the QoE of watching that specific item.

The subjects were shown the $60$ items in a random order, in full-screen mode. Randomisation is important in order to avoid creating a bias on the participants. The evaluation took approximately one hour for each candidate. After performing the experiments, the obtained opinion scores were averaged to obtained a single MOS for each item, which was used for the training of the CNNs. Moreover, further processing was applied to the QoE scores in terms of discretisation, by grouping the average QoE values into classes.

\section{Proposed method} \label{Fourth_Section}
In this section a description of the proposed methodology for predicting VQA of video signals is presented. Details on how the CNN used for performing the VQA estimation was designed are also presented. 

\subsection{Data pre-processing and augmentation}
CNNs require a considerable amount of data in order to ensure that sufficient generalisation is achieved during the training. Moreover, directly processing frames of $1920 \times 1080$ samples is too complex from the perspective of memory requirements during the training, as well as during the classification when applying the CNNs for performing the VQA prediction.

In order to tackle this problem, a patch-based data augmentation technique was applied to the training data. For simplicity, only the luminance component was considered in the proposed approach. Each video volume (corresponding to the three dimensional matrix of luminance pixel values of size $1920 \times 1080 \times N$, where $N$ is the number of frames in the sequence) was split into a sequence of non-overlapping cubic patches of size $k \times k \times k$ containing pixel values, which serve as the examples to be used for training the CNN. Since each entire video sequence is annotated with a single label, the same VQA label is propagated to each of the training samples belonging to the dataset item. The obtained training examples are then grouped together in order to form the dataset which is used for training the CNN.

\subsection{Spatio-temporal learning with higher-order CNNs}
The proposed method fits within classification-flavoured supervised learning, employing higher-order CNNs for efficient spatio-temporal feature learning of the video content. Unlike conventional neural networks, CNNs employ the notion of local receptive fields in order to effectively extract features from raw data. More specifically, each locally connected input subset of the input neuron is mapped to a single output neuron, a process which is performed in a stacked manner throughout convolutional layers, in order to capture as many representative features as possible. The connection between input and output neurons is performed via convolutions by means of trainable kernels, namely filters with specific filter coefficients. The number of such filters can be  trimmed using pooling layers in order to avoid over-fitting issues. 

Formally, the value of a convolved output neuron at position $(k,l)$ in the j-th feature map of the i-th layer can be expressed as follows:

\begin{equation} \label{Fifth_Equation}
y_{i,j}^{k,l}=f(\sum_{i=0}^{H-1} \sum_{j=0}^{W-1} w_{i,j}x_{(k+i)(l+j)}+b_{i,j}),
\end{equation}
where $f(.)$ is an activation function, $w_{i,j}$ stands for the value of the kernel connected to the current feature map at the position $(i,j)$, $x_{(k+i)(l+j)}$ represents the value of the input neuron, $b_{i,j}$ is the bias of the computed feature map, and $H$ and $W$ are the height and width of the kernel respectively. 

Processing video signals implies that in addition to spatial information, temporal (inter-frame) redundancies also exist among neighbouring frames. In order to exploit such information, a 3D-CNN approach is proposed in this paper for effective preservation of temporal and motion features which may be essential to VQA. 3D convolution is an extension of the 2D convolution operation,  in which the learnable convolution kernel is a $3$-dimensional cube which considers local spatial regions extracted from adjacent frames. Formally, the third-order analogue of Equation \ref{Fifth_Equation} can be expressed as follows:
\begin{equation} \label{Sixth_Equation}
y_{i,j,p}^{klm}=f(\sum_{i=0}^{H-1} \sum_{j=1}^{W-1} \sum_{p=1}^{T-1} w_{i,j,p}x_{(k+i)(l+j)(m+p)}+b_{i,j,p}),
\end{equation}
where $T$ depicts the temporal dimension of the kernel cube, and the respective quantities in Equation \ref{Fifth_Equation} are extended to their three dimensional counterparts. Adopting such a strategy, the feature maps in the convolution layer are connected to multiple contiguous frames in the previous layer, leading to better capturing motion information.

\begin{figure}[t]
\centering
\includegraphics[width=250pt]{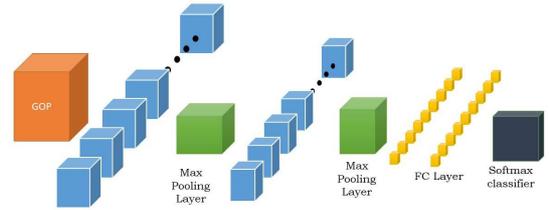}
\caption{Proposed CNN-based architecture for QoE prediction.}
\label{Proposed_Method_QoE_Prediction_Modeling_First_Figure}
\end{figure}

\subsection{QoE prediction modelling }
The architecture of the CNN used throughout this paper was inspired by work in human action recognition \cite{tran2015learning}. The network utilises multiple stacks of convolutional layers, max-pooling layers, normalisation layers and Rectified Linear Unit (ReLU) activation layers. It has been shown (for instance in \cite{tran2015learning}) that small receptive fields of $3\times3$ convolution kernels may lead to higher classification accuracies than using larger kernels. Extending the assumption to the temporal dimension, while also noting that temporal redundancy generally quickly decreases with time, an equivalent temporal length of $3$ was also considered in this paper. A representation of the architecture of the network described in this section is presented in  Figure \ref{Proposed_Method_QoE_Prediction_Modeling_First_Figure}.

In terms of the number of stacks of layers employed, architectures comprising of $2$ and $3$ layers were explored. In the proposed model, each convolution layer is always followed by a max-pooling layer. These layers are then followed by two fully-connected layers. Finally, a final soft-max layer is considered for the prediction task. 

Regarding the number of filters, a "doubling-depth" strategy was adopted in which deeper layers in the network utilise double the number of  filters used for convolution in the previous layer. This approach was shown to be successful in tasks in which it is crucial for the network to learn abstract features \cite{tran2015learning}.

In order to preserve as much temporal information as possible, the max-pooling layer following the first convolutional layer in the network has a size of $2\times2\times1$ (corresponding to reducing the input signal by a factor of $4$). This ensures that temporal redundancies are not discarded in the initial phases of the signal path. All subsequent max-pooling layers following deeper convolutional layers in the network have a size of $2\times2\times2$ samples, which corresponds to reducing the input signal by a factor of $8$. The first and the second fully-connected layer have $1024$ and $512$ outputs, respectively. 

In terms of the optimiser, both standard SGD optimisation with learning rates of $10^{-2}$ and $10^{-3}$ , as well as Adagrad optimisation \cite{duchi2011adaptive} were investigated. The latter was selected due to its adaptive selection of reducing the learning rate of parameters with high gradients and, conversely, increasing that of parameters which have small or infrequent updates. Categorical Cross-Entropy was used as the loss function. 

Training was performed using varying number of epochs (e.g. $50$, $500$, $750$), with the goal of observing how the prediction accuracy changes by increasing the epochs. The CNN performance was measured in terms of Classification Accuracy and Loss with respect to the total number of epochs, measured for each patch. However, patches are extracted from specific video sequences, and it is therefore important to also assess the performance of the network in terms of its ability in predicting the VQA of an entire video sequence. To that extent, an aggregation process was considered in order to obtain a single accuracy score for each item in the dataset. At this purpose different strategies can be utilised. In this paper, the two approaches  described in \cite{kim2017deep} are considered, namely:

\begin{itemize}
\item Aggregation in terms of majority-vote strategy;
\item Using a pre-trained model strategy.
\end{itemize}

\begin{figure}[t]
\centering
\includegraphics[width=250pt]{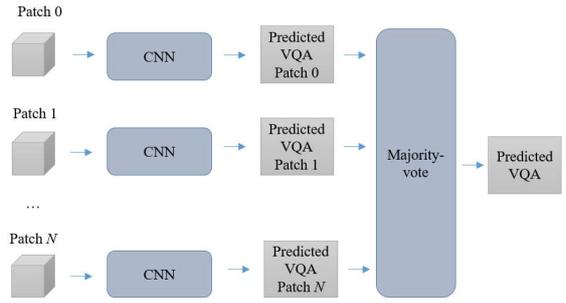}
\caption{Majority-voting strategy for patch aggregation.}
\label{Proposed_Method_QoE_Prediction_Modeling_Second_Figure}
\end{figure}

When considering aggregation in terms of majority-vote strategy, patch aggregation is not performed during the actual training process. On the contrary, training is performed as if each patch is independent from the others disregarding the fact that patches may be extracted from the same item. After training, during the classification, the labels associated with patches belonging to the same item are grouped together, and the label that is selected most frequently is selected as label for the entire video sequence. This strategy is represented in Figure \ref{Proposed_Method_QoE_Prediction_Modeling_Second_Figure}. When adopting this strategy, each patch is therefore independently "regressed" onto the global subjective score for the video sequence.

\begin{figure}[t]
\centering
\includegraphics[width=250pt]{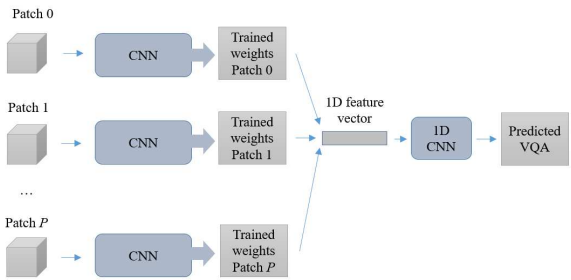}
\caption{Patch aggregation using pre-trained model.}
\label{Proposed_Method_QoE_Prediction_Modeling_Third_Figure}
\end{figure}

As an alternative to the majority-vote strategy, a pre-trained model can instead be considered. In this case the patch aggregation process is incorporated within the training process, as illustrated in Figure \ref{Proposed_Method_QoE_Prediction_Modeling_Third_Figure}. The aforementioned 3D-CNN is initially trained on video signal data using each patch individually, as is the case for the majority-vote strategy. The network gains knowledge from this data and embeds such knowledge within the trained weights. These weights can be extracted and then transferred to another CNN, to perform the actual classification. Formally, a given number of patches are randomly extracted from a given video. Refer to each patch as $p = 0, 1, ... P-1$ where $P$ is the total number of extracted patches (a value of $P=1000$ was used in the rest of this paper). After each patch is fed to the CNN, the training is performed by means of backpropagation. The trained set of weights is then extracted and arranged in a one dimensional vector $W_p$ of length $T$, where $T$ is the total number of filter coefficients in the CNN. Finally, these vectors are then arranged in a single feature vector $W = [W_0, W_1, ..., W_P-1]$ of length $P \cdot T$.

These feature vectors, extracted from the set of training videos, are then used to train an additional one-dimen-sional CNN. The input to the CNN is a one-dimensional vector of length $P \cdot T$.  The structure of this CNN is identical to that of the 3D CNN, but the 3D filters are replaced by their 1D counterparts. The output is again formulated in terms of classification as a single discrete value, representing the predicted VQA of the current video. 

When applying this approach on a new video from a test set, the first 3D CNN is trained again (using the $P$ patches extracted from the test video), to obtain the feature vector $W$. This is then input to the second CNN, to output the final VQA prediction. Intuitively, this strategy implies that instead of training the 1D CNN from scratch, the learned features of each video item are transferred from the 3D CNN. This allows good classification accuracy even if the second 1D CNN is actually trained with a very limited number of training samples (equal to the number of video sequences in the training set). This strategy overcomes the limitations of the data augmentation technique used, while avoiding usage of the majority-vote strategy which may lead to unsatisfactory results. A detailed analysis of the performance of both strategies is presented in the rest of this paper.

To further quantify the performance of the proposed system, apart from the Classification Accuracy and Loss, other metrics were also used, namely True Positive Rate (TPR), False Negative Rate (FNR), False Positive Rate (FPR),  True Negative Rate (TNR) and accuracy-per-class.

\section{Experimental evaluation}
\label{Fifth_Section}
In this section, results of the adopted CNNs under different scenarios using various parameters are presented and discussed, to illustrate the performance of the methodologies introduced in this paper.

\subsection{Effects of compression in VQA prediction}
In a first set of experiments, the proposed DL-based VQA prediction was assessed only taking into account the effects of compression. To this end, network conditions were considered constant. The experiments only utilise $15$ items in the dataset, corresponding to the $3$ UGC clips compressed with the $5$ QP values, transmitted using the best available network conditions. In all experiments presented in this subsection, the discrete labels used for training and testing were obtained quantising the average MOS ratings using non-overlapping intervals of equal sizes (of size $1.33$), resulting in three labels, where label 1 corresponds to the lowest annotated QoE and label 3 corresponds to the highest QoE.

The training samples used to feed the CNN are patches of size $16 \times 16 \times 16$. For the purpose of this experiment, a two-layers model was used, with the parameter specifications described in Section \ref{Fourth_Section}. The effects of using different optimisers as well as varying number of trainable filters were investigated. Also, increasing number of epochs was used for training, in order to highlight the effects of the number of epochs on the network performance.

\begin{figure}[!ht]
\centering
\includegraphics[width=120pt]{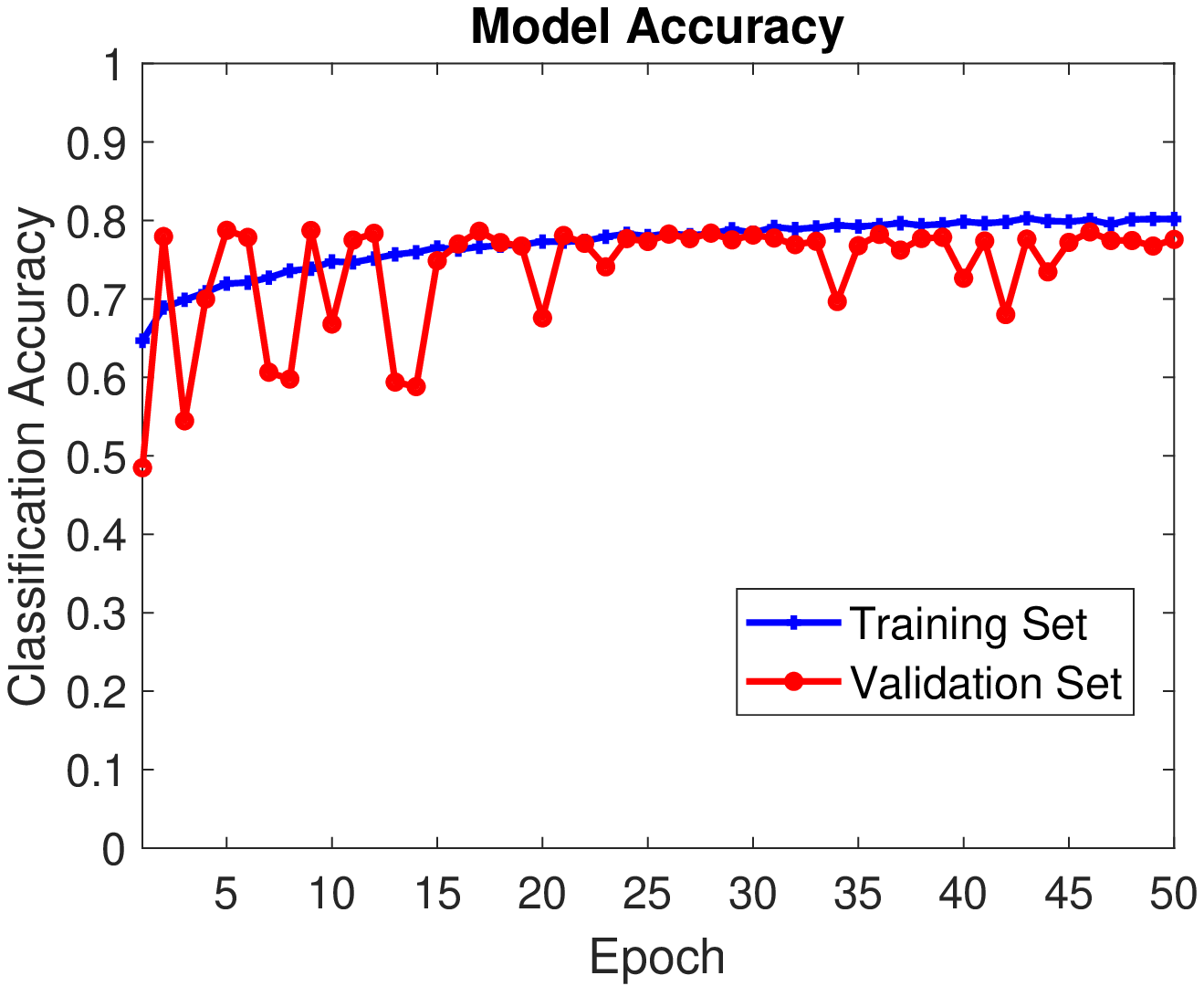}
\includegraphics[width=120pt]{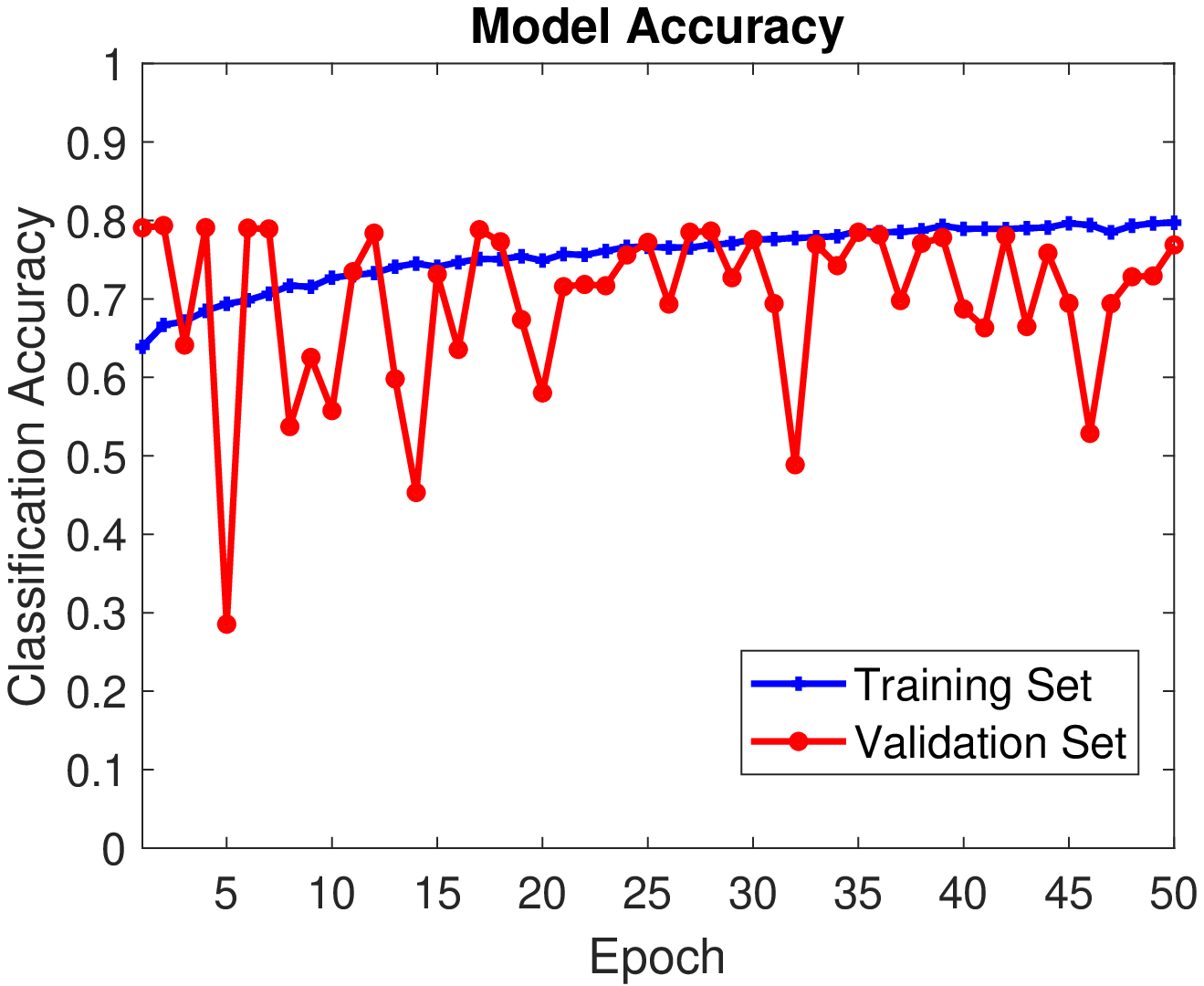}
\caption{Patch-wise classification accuracy for the Adagrad (left) and SGD (right) optimisers with 16-32 filters.}
\label{Results_First_Experiment_First_Figure}
\end{figure}

First, a two-layered CNN was considered, with $16$ filters in the first layer and $32$ filters in the second layer. A comparison between the conventional SGD optimiser and the Adagrad optimiser is first performed using 50 epochs for training. Results of this analysis are shown in the plots in Figure \ref{Results_First_Experiment_First_Figure}. The plot shows the patch-wise CNN accuracy for training and validation with respect to the number of epochs up to which the network was trained.

The results of this experiment show that the SGD optimiser produces less stable results than the Adagrad optimiser. Both optimisation algorithms lead to validation accuracy of up to $70\%$. In the case of using Adagrad, the model accuracy steadily increases with the epoch, and the accuracy with the validation set stabilises towards that of the training set, showing that the model is capable in generalising well during the training. 

\begin{figure}[!ht]
\centering
\includegraphics[width=120pt]{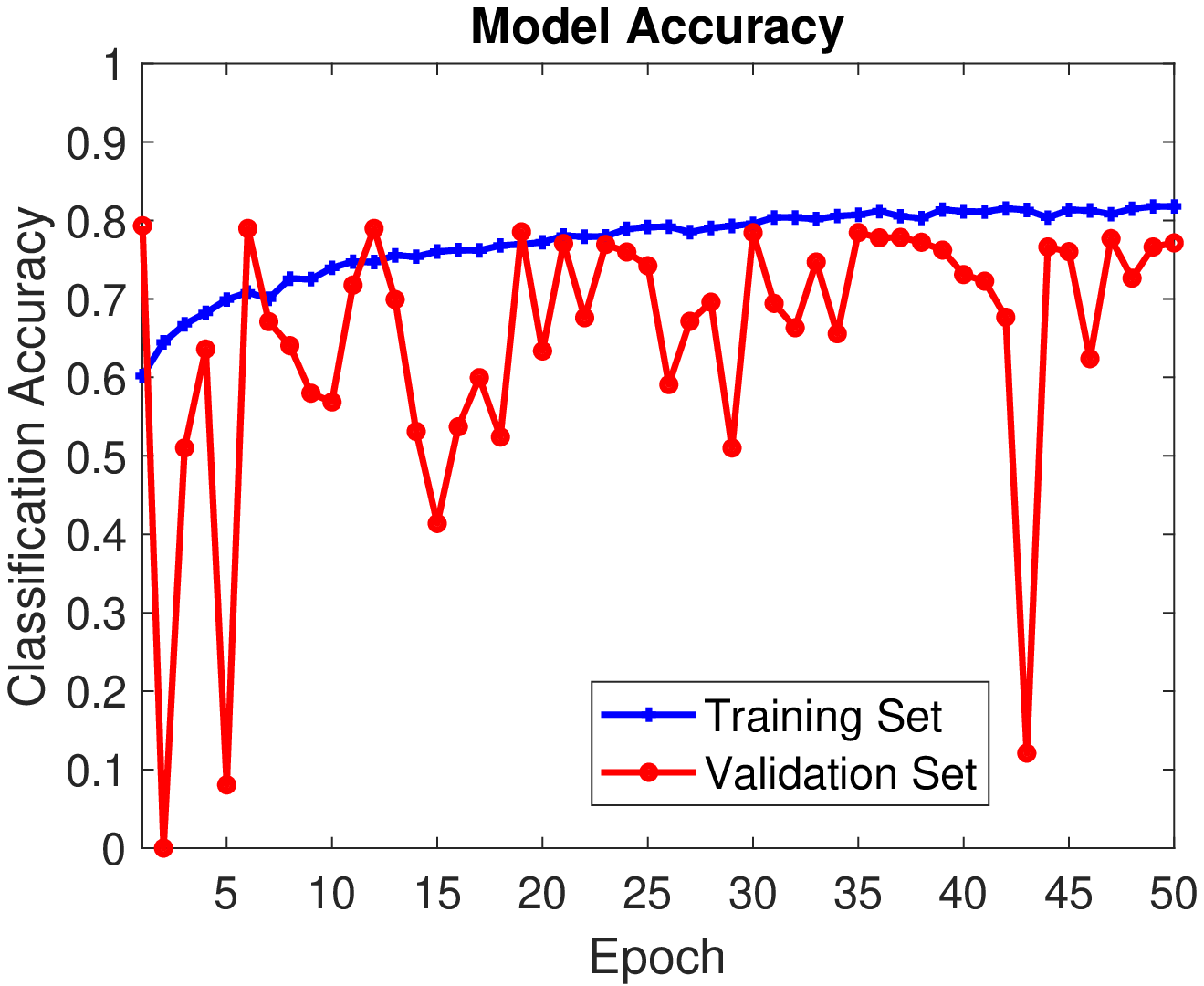}
\vspace{10pt}
\includegraphics[width=120pt]{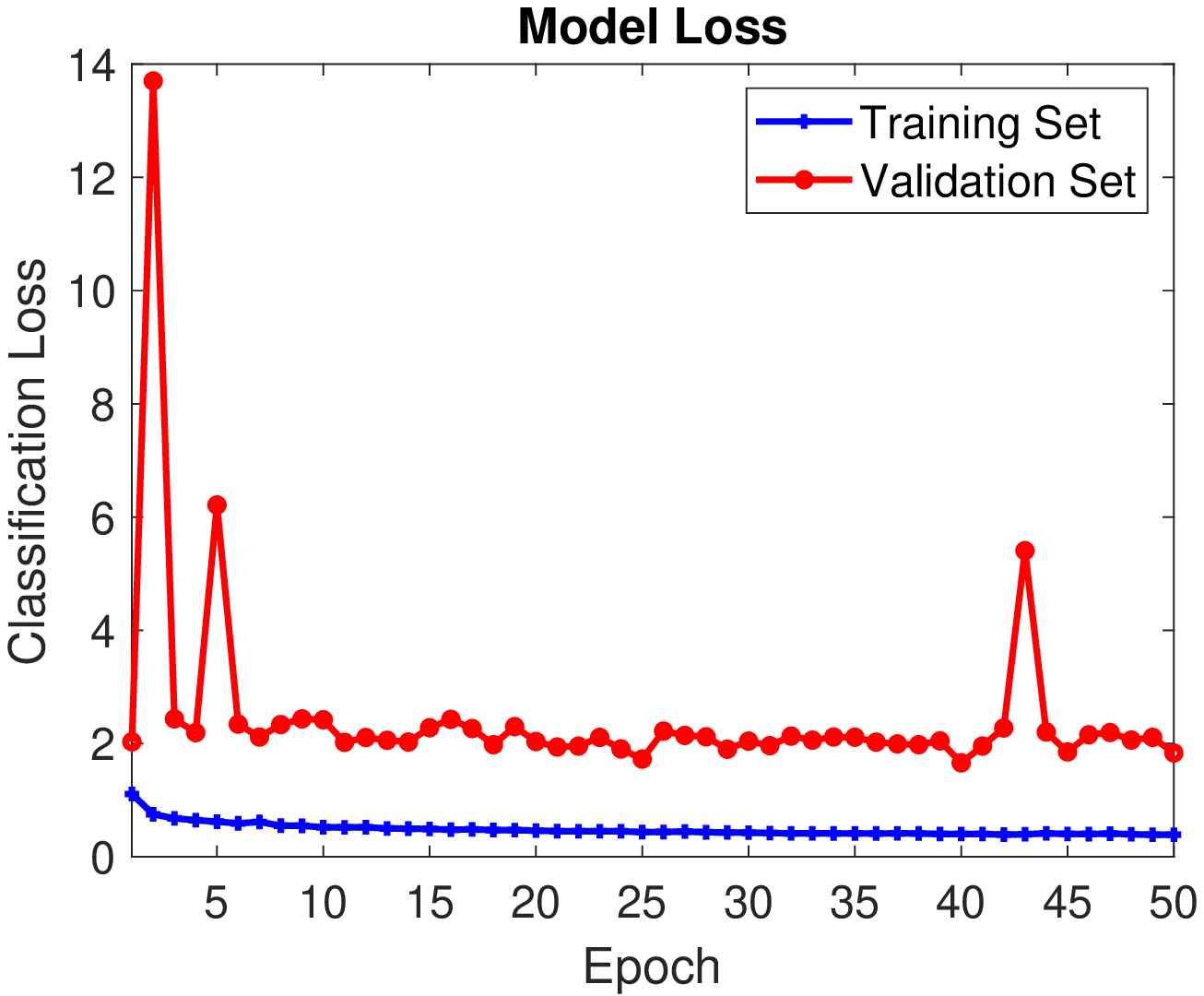}
\vspace{-10pt}
\caption{Patch-wise classification accuracy (right) and loss (left) for SGD optimiser with 128-256 filters.}
\label{Results_First_Experiment_Second_Figure}
\end{figure}

In an attempt to obtain more representative features, the number of trainable filters per layer was also investigated. In order to isolate these effects, the conventional SGD optimiser was again used for these experiments. The number of filters was increased by a factor of $4$, corresponding to using $64$ filters in the first layer and $128$ filters in the second layer. The performance of the system in terms of model accuracy as well as model loss is illustrated in Figure \ref{Results_First_Experiment_Second_Figure}.

The plots in Figure \ref{Results_First_Experiment_Second_Figure} (left) show that the CNN achieves good classification accuracy, even though the accuracy for specific patches is considerably lower than what obtained with smaller number of filters, pointing to the fact that the generalisation properties of the CNN may have decreased as an effect of increasing the number of filters. The decaying trend of the model loss depicted in Figure \ref{Results_First_Experiment_Second_Figure} (right) further confirms this claim.  

\begin{figure}[!t]
\centering
\includegraphics[width=120pt]{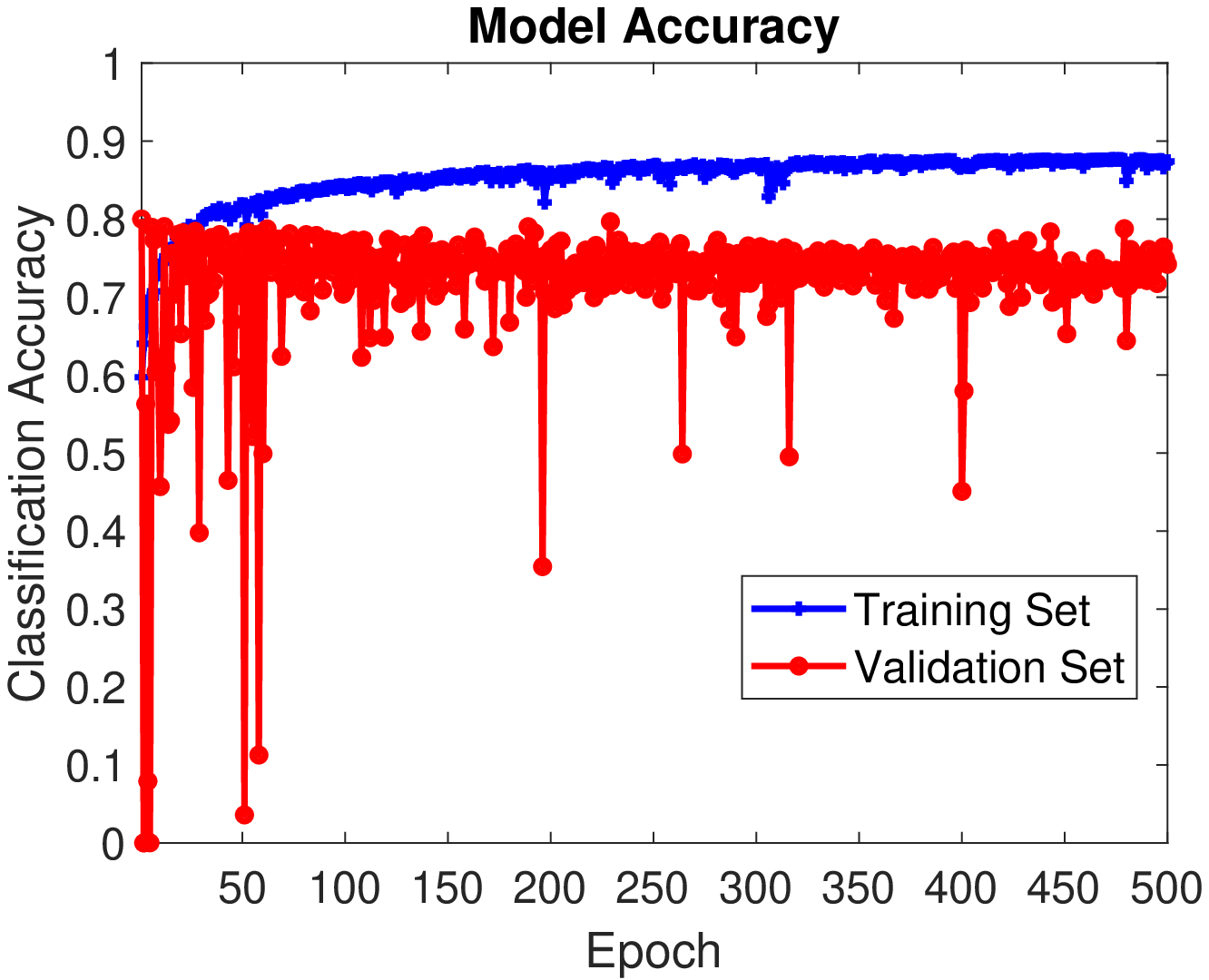}
\vspace{10pt}
\includegraphics[width=120pt]{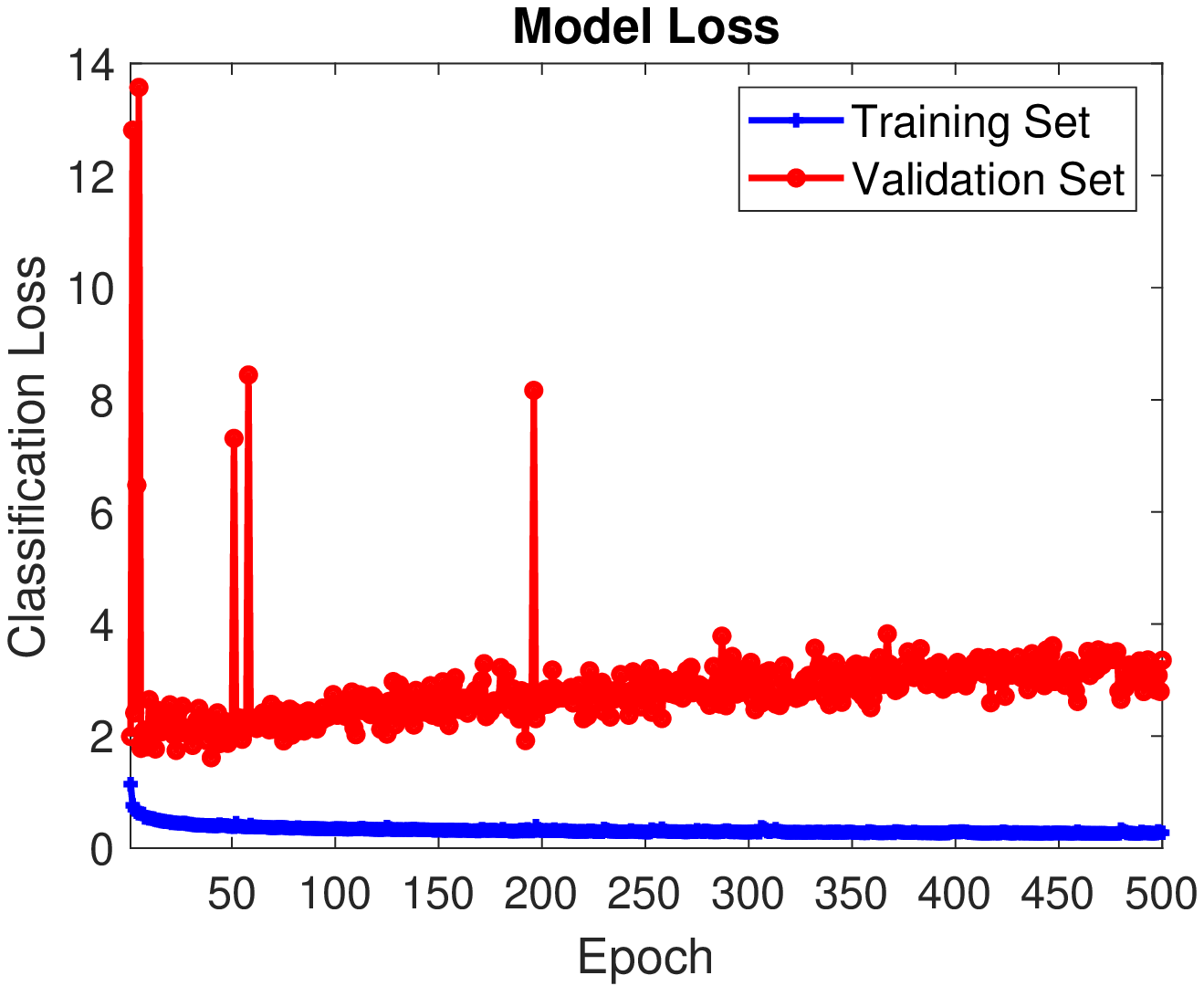}
\caption{Patch-wise classification accuracy (left) and loss (right) for training the system with 500 epochs.}
\vspace{-5pt}
\label{Results_First_Experiment_Third_Figure}
\end{figure}

To increase the generalisation properties, an additional experiment was then performed in which training was performed with $500$ epochs, instead of $50$ epochs as in the aforementioned experiments. Results of this experiment are presented in Figure \ref{Results_First_Experiment_Third_Figure}.

It can be observed that while the accuracy stabilises and does not seem to improve further with the increasing number of epochs, the system increases its generalisation properties, resulting in a decreasing number of patches with low classification accuracy. This is also reflected in less peaks in the loss function shown on the right of Figure \ref{Results_First_Experiment_Third_Figure}.

Table \ref{Results_First_Experiment_First_Table} reports the classification accuracy of the system when utilising the majority-vote strategy to obtain item-wise classification accuracy (as opposite to the patch-wise results shown in the aforementioned plots). Sequence-wise accuracy of $80\%$ is obtained in the experiments. Due to the generally good accuracies obtained patch-wise, when utilising a majority-vote strategy, the CNN obtains consistently good results, smoothing out the effects of the varying conditions presented in these experiments.
\begin{table}[]
\vspace{15pt}
\footnotesize
\centering
\caption{Patch-wise (P) and Sequence-wise (S) classification accuracy for various optimisers, number of filters and epochs. }
\label{Results_First_Experiment_First_Table}
\begin{tabular}{|c|c|c|c|c|}
\hline
\textbf{Optimiser} & \textbf{Filters} & \textbf{Epochs} & \textbf{P-Accuracy} & \textbf{S-Accuracy} \\ \hline
Adagrad & 16-32 & 50 & 77.58\% & 80\% \\ \hline
SGD & 16-32 & 50 & 76.88\% & 80\% \\ \hline
SGD & 128-256 & 50 & 77.14\% & 80\% \\ \hline
SGD & 128-256 & 500 & 74.20\% & 80\% \\ \hline
\end{tabular}
\end{table}

\subsection{Joint effects of compression and network conditions in VQA prediction}
In a second set of experiments, the analysis was extended to include the rest of the dataset, considering impairments due to both compression and transmission. A similar analysis as that from the previous subsection was performed in order to investigate the influence of the parameters on the CNN performance. In the initial set of experiments presented in this subsection, non-overlapping intervals of equal size $1.33$ were used for discretisation, resulting in three labels.

\begin{figure}[!t]
\centering
\includegraphics[width=120pt]{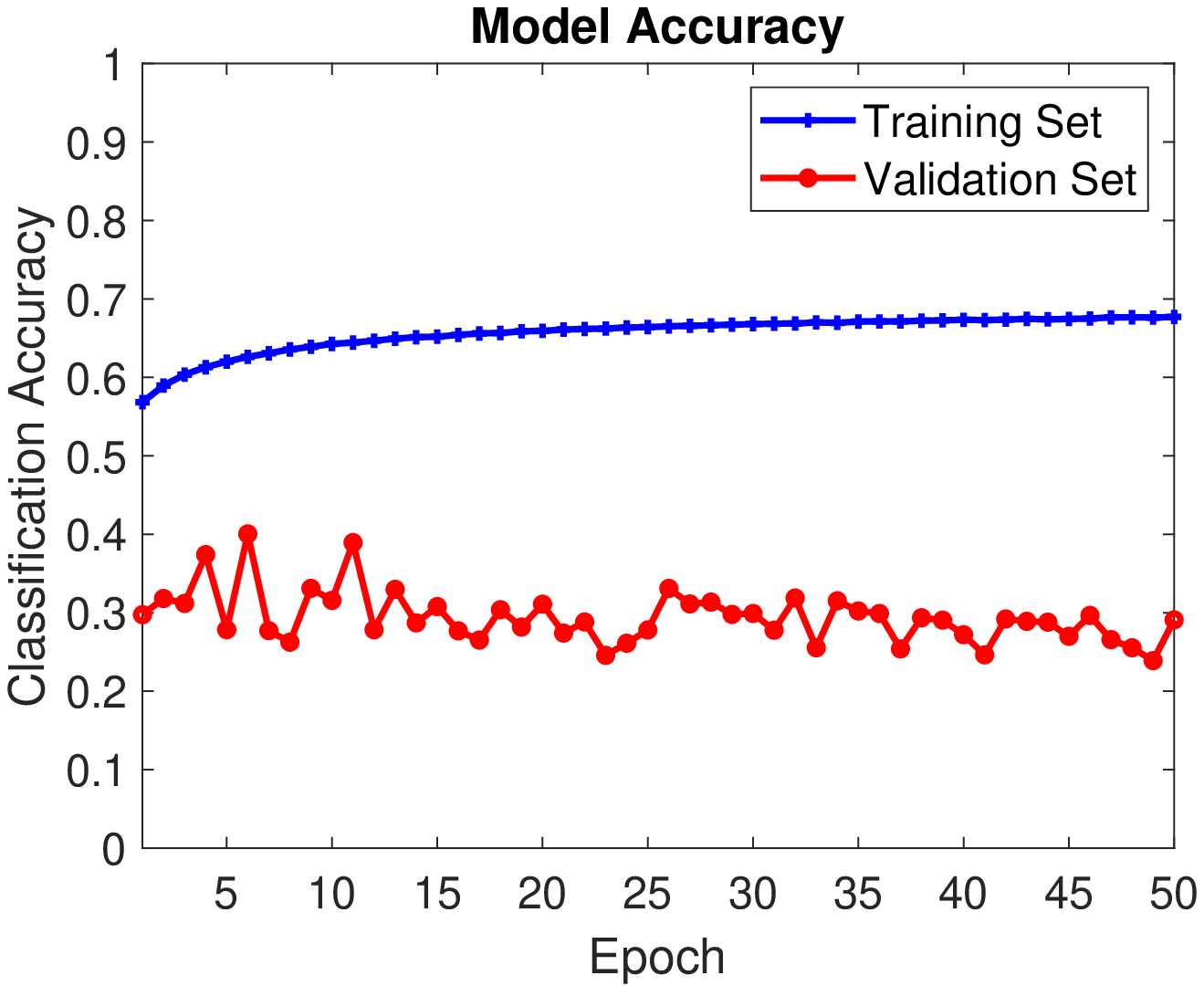}
\includegraphics[width=120pt]{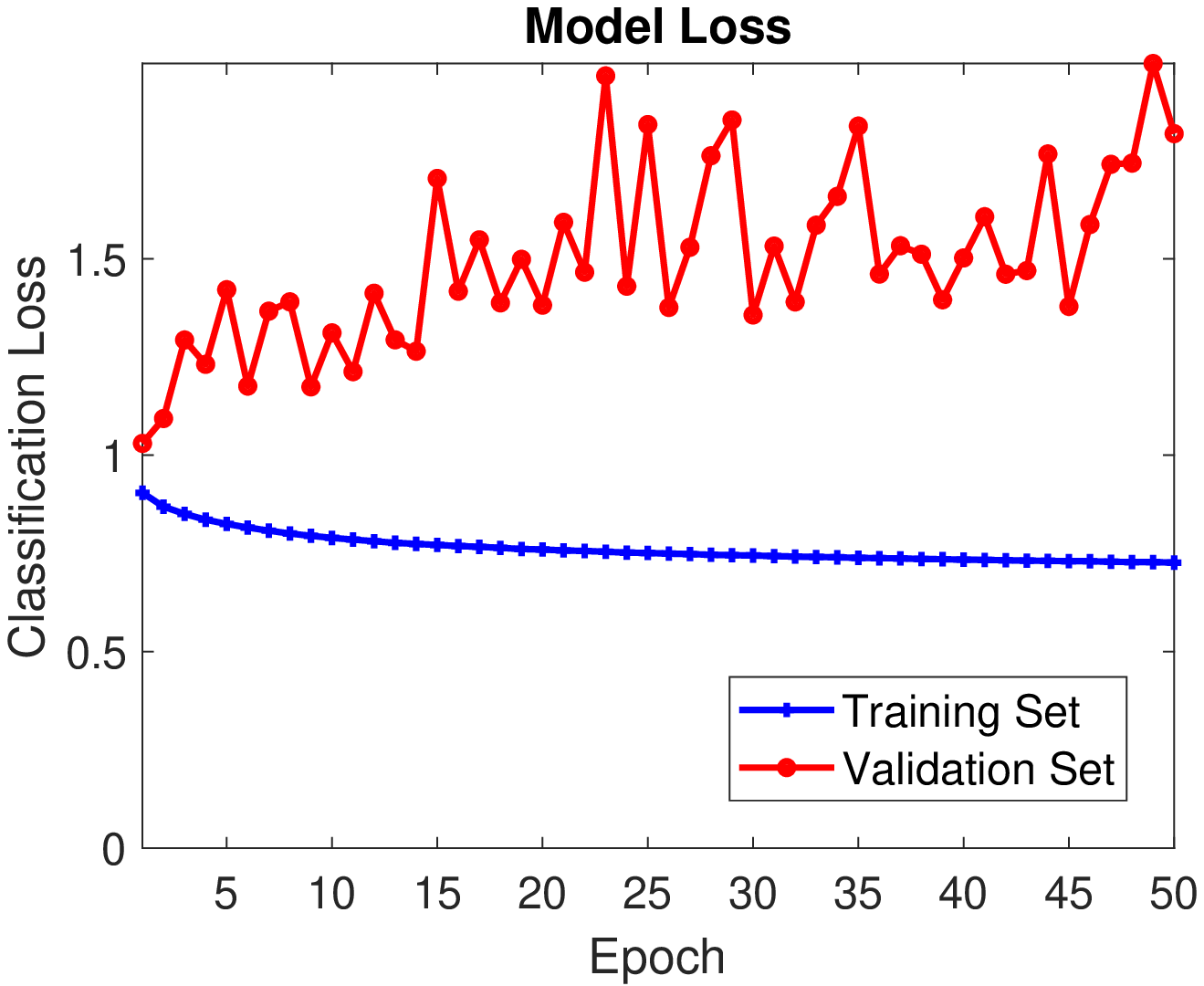}
\caption{Patch-wise classification accuracy (left) and loss (right) for SGD optimiser with 16-32 filters and 50 epochs.}
\label{Results_New_First}
\end{figure}

First, the conventional SGD optimiser is used, using the CNN with $16$ filters in the first layer and $32$ filters in the subsequent layers. The results obtained in terms of patch-wise accuracy and loss are depicted in Figure \ref{Results_New_First}.As can be seen, the network does not generalise well, providing low classification accuracy on the validation set. The model loss increases with the number of epochs, highlighting that the CNN parameters are not suitable to model the combined effects of network and compression.

\begin{figure}[!b]
\centering
\includegraphics[width=120pt]{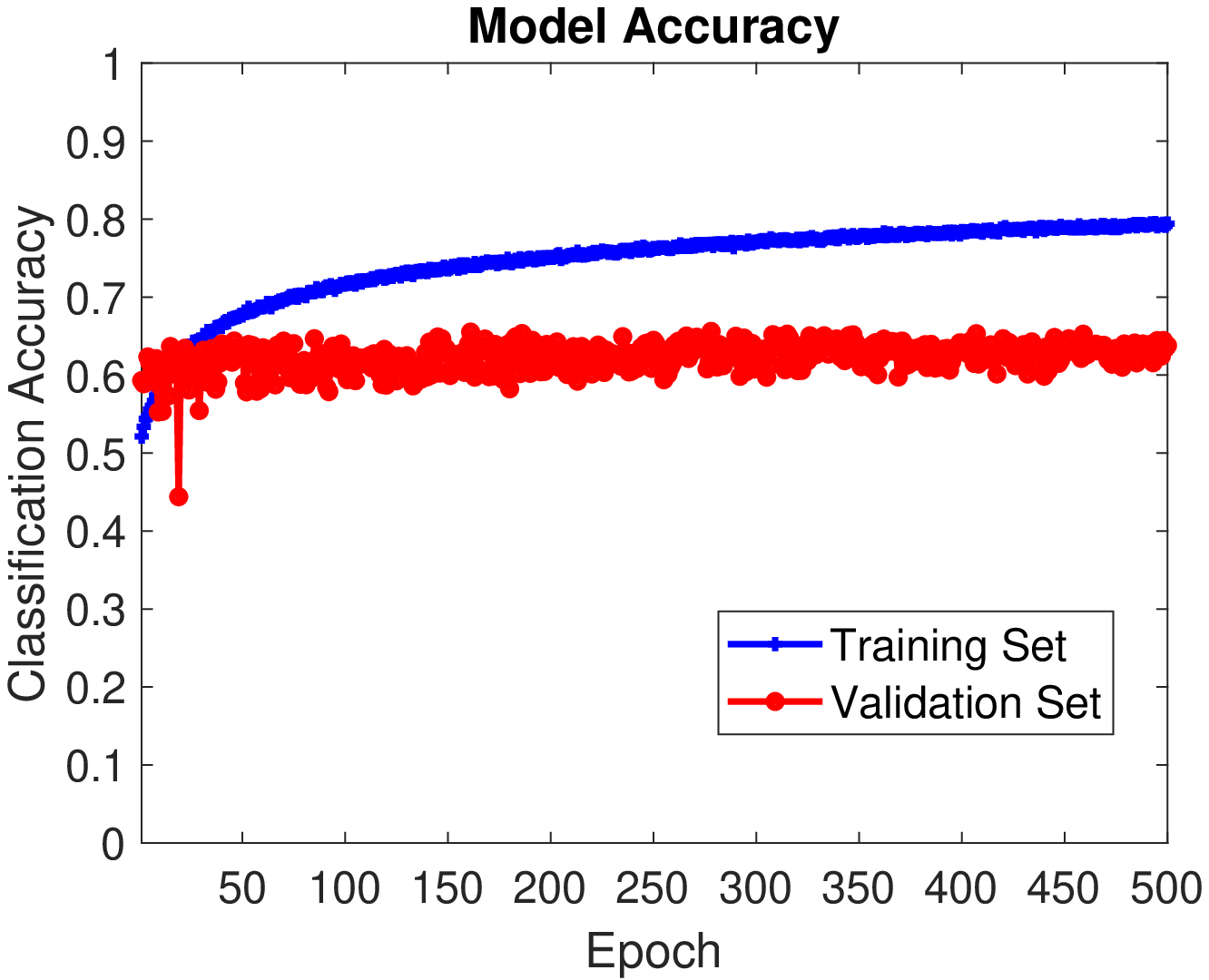}
\vspace{10pt}
\includegraphics[width=120ptt]{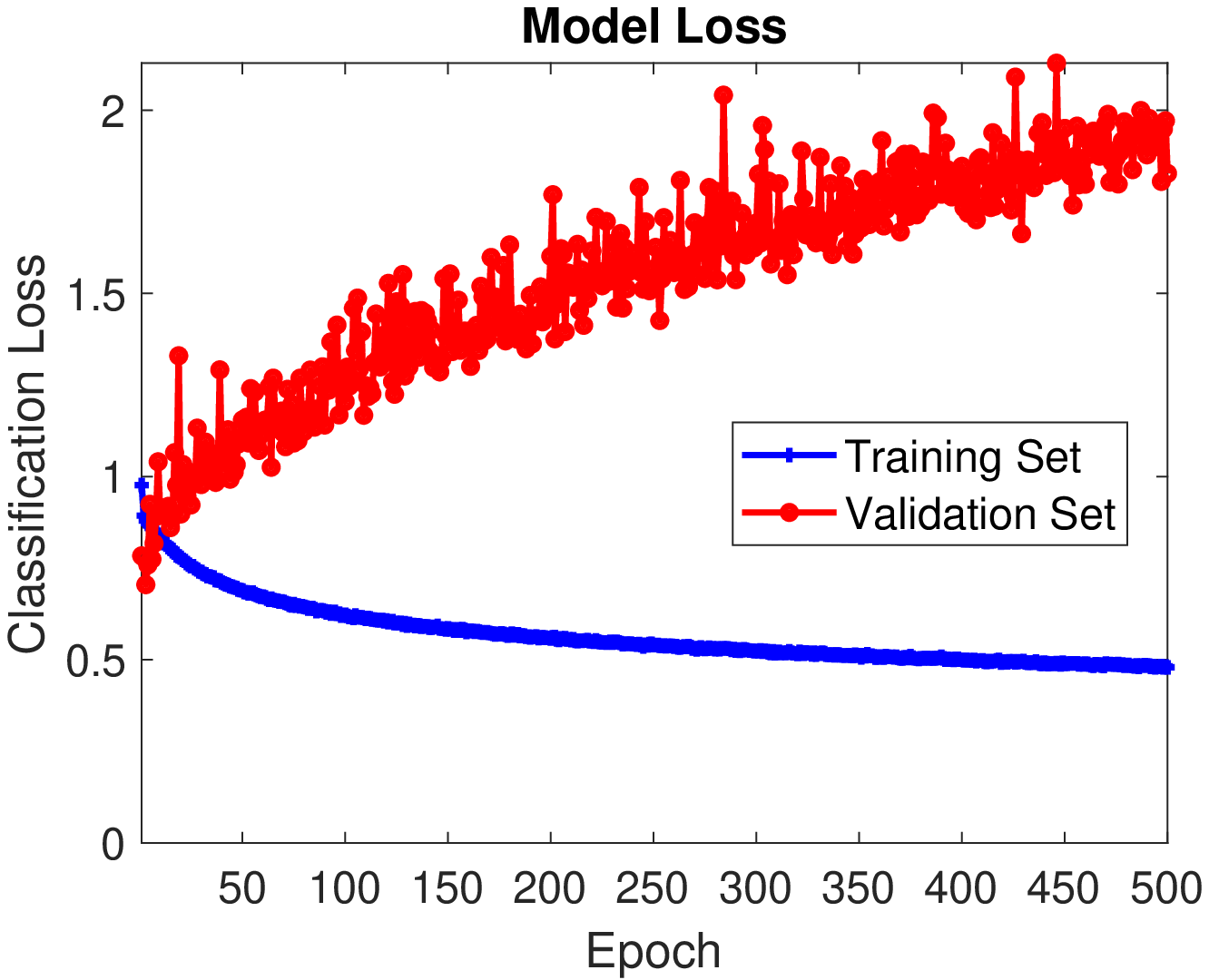}
\caption{Patch-wise classification accuracy (left) and loss (right) for Adagrad optimiser with 16-32 filters and 500 epochs.}
\vspace{-5pt}
\label{Results_New_Second}
\end{figure}

Given that the Adagrad optimiser was shown to perform better in the results presented in the previous subsection, this optimiser was also tested under these conditions. Moreover, increasing the number of epochs was also shown to have a positive effect on the CNN performance and therefore, $500$ epochs were used for the subsequent test. Under these  conditions, the CNN performances increase considerably, especially in terms of validation accuracy, as shown in Figure \ref{Results_New_Second}.

The results of the experiment demonstrate that Adagrad behaves again more consistently than SGD in terms of patch-wise classification. These two models were compared also using the majority-vote strategy to obtain sequence-wise accuracy. The first model utilising the conventional SGD optimiser resulted in an accuracy of $66.67\%$ , whereas the second model utilising the Adagrad optimiser outperformed that, reaching an accuracy of $73.33\%$.

\begin{figure}[!b]
\centering
\includegraphics[width=120pt]{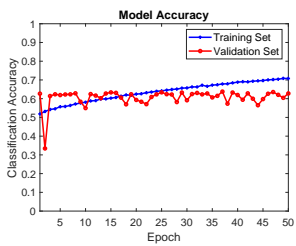}
\caption{Patch-wise classification accuracy using 128-256 filters and 50 epochs.}
\vspace{-5pt}
\label{Results_New_Third}
\end{figure}

In an attempt to obtain better classification performance, the number of filters per layer was increased to $128$ in the first layer and $256$ for the subsequent layers. The obtained patch-wise classification accuracy is shown in Figure \ref{Results_New_Third}.

\begin{figure}[!t]
\centering
\includegraphics[width=120pt]{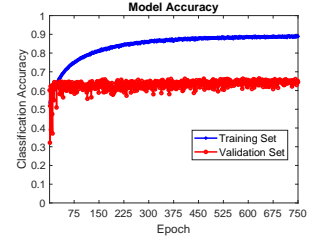}
\caption{Patch-wise classification accuracy obtained with final model parameters, using majority-vote strategy. }
\vspace{-5pt}
\label{Results_Third_Experiment_First_Figure}
\end{figure}

The aforementioned results were used as basis to construct a final model for the CNN, using $128$ filters in the first layer and $256$ in the subsequent layers, trained with a large number of epochs, and making use of the SGD optimiser. Figure \ref{Results_Third_Experiment_First_Figure} presents results of this experiment, ending up with patch-wise accuracy of up to $64.43$\% and sequence-wise accuracy of $80\%$.  

\begin{figure}[!ht]
\centering
\includegraphics[width=140pt]{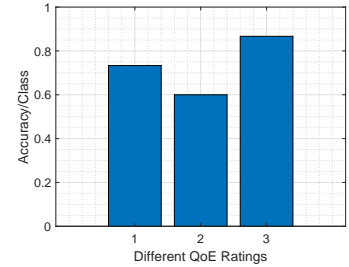}
\caption{Classification accuracy per label.}
\label{Results_Second_Experiment_First_Figure}
\end{figure}

An analysis of the label-wise accuracy of classification was performed, to understand whether the CNN is better in predicting specific labels.  Figure \ref{Results_Second_Experiment_First_Figure} shows the accuracy of classification obtained for each label. As expected, the labels at the extremes are easier to predict (in that the corresponding impairments are either very clear, or the items provide very high QoE). Nonetheless, the CNN predicts each label with an accuracy of at least $60\%$.

\begin{figure}[!ht]
\centering
\includegraphics[width=160pt]{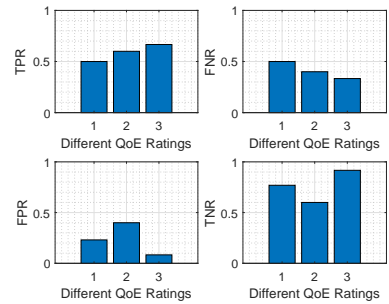}
\caption{Statistical Hypothesis Testing performance metrics. }
\label{Results_Second_Experiment_Second_Figure}
\end{figure}

To further analyse the performance of the CNN label-wise, the classification problem was treated as a statistical hypothesis testing, and its performance was investigated in terms of respective commonly used performance metrics such as the aforementioned TPR, FNR, FPR and TNR. These results can be seen in Figure \ref{Results_Second_Experiment_Second_Figure}. It is commonly accepted that FPR below $25\%$ for two out of three classes indicates good classification capabilities.

\begin{figure}[!t]
\centering
\includegraphics[width=130pt]{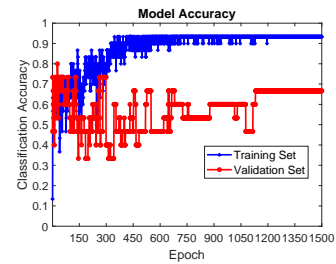}
\caption{Patch-wise classification accuracy obtained with final model parameters, using the pre-trained model strategy. }
\vspace{-5pt}
\label{Results_Pretrained_Experiment_First_Figure}
\end{figure}

Finally, the model was tested using the pre-trained strategy for patch aggregation. A two-layered 3D CNN with $128$ filters in the first layer and $256$ filters in the subsequent layers was used. This was followed by a 1D CNN, consisting of $3$ layers with $64$, $128$ and $256$ filters per layer, respectively. When using this strategy, the sequence-based classification is incorporated during the training. Still, even though the training of the two subsequent CNNs is performed with the goal of obtaining a single sequence-wise classification, the performance of the first (3D) CNN can still be assessed in order to evaluate its classification accuracy. Results of this test are shown in Figure \ref{Results_Pretrained_Experiment_First_Figure}, where it can be seen that good classification is obtained patch wise, even though the results in the validation set are less stable than those obtained using the majority-vote strategy. Sequence-wise, an accuracy of $66.7\%$ was obtained. This shows that the pre-trained model does not perform as well as the model making use of the majority-vote. Nonetheless, using such model avoids the need for grouping the labels and performing majority-vote patch aggregation, because a single classification can directly be obtained as output of the 1D CNN. Due to the fact that better performance are obtained with the majority-vote strategy, this was selected as the model of choice, unless the application requires the network to directly provide a single classification for each video item.

Different discretisation strategies can be adopted when transforming the continuous average MOS ratings into discrete labels. An analysis of different strategies was performed in order to verify which strategy to adopt when constructing the discrete classes. Four different scenarios are considered, corresponding to dividing the continuous range of MOS ratings (from $1$ to $5$) in non-overlapping intervals using different sizes of $1.33$, $0.5$, $0.25$ or $0.125$. A total of $3$, $5$, $8$ and $14$ labels are obtained in each case, respectively.

\begin{figure}[!t]
\centering
\includegraphics[width=140pt]{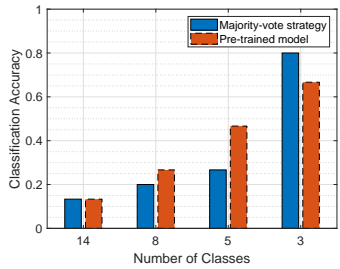}
\caption{Sequence-wise classification accuracy with majority-vote strategy (blue) and pre-trained model with different number of output labels (classes).} 
\label{Results_Third_Experiment_Second_Figure}
\end{figure}

Results of these experiments were computed in terms of sequence-wise accuracy, using both patch aggregation strategies described in this paper. Results can be seen in Figure \ref{Results_Third_Experiment_Second_Figure}. It can be observed that the classification accuracy drops considerably when considering smaller sizes of the discretisation intervals. This is to be expected, in that the problem becomes more complex given as the CNN has to classify each item into more classes. Interestingly, while the majority-vote strategy outperforms the pre-trained model in the case of $3$ labels, opposite behaviour is obtained in other cases, where the pre-trained model performs better. Especially in the case of $5$ labels, the pre-trained model achieves a much higher accuracy than the majority-vote strategy. Nonetheless, the results justify the choice of using a total of $3$ output labels, which outperforms all other cases.

The results presented in this section justify the choice of the model parameters, and highlight an acceptable performance of the proposed approach stressing its good generalisation properties, especially taking into account the complexity of the problem tackled within this paper.

\section{Conclusions} \label{Sixth_Section}
This paper presents a novel DL approach to predict the perceived quality of video content affected by compression and network conditions.The approach is based on higher-order CNNs, for learning efficient spatio-temporal feature representations. The method was designed by means of an ad-hoc dataset comprising various pieces of challenging UGC material affected by HEVC encoding and various network conditions. The problem was posed as a classification problem, where two strategies are proposed in order to obtain sequence-wise classification. Extensive evaluation is provided to identify suitable network parameters, showing that the method is capable of achieving consistent classification accuracy under challenging conditions.

\section{Acknowledgments}
The work leading to this paper was co-supported by the Greek General Secretariat for Research and Technology (GSRT), the Hellenic Foundation for Research and Innovation (HFRI) and by the project COGNITUS, which received funding from the European Union's Horizon 2020 research and innovation programme under grant agreement No 687605.

\section*{References}

\bibliography{Signal_Processing_Image_Communication_2018_Paper_References}

\end{document}